\documentclass[useAMS,usenatbib]{mn2e}

\newcommand{\B}{Bayes\textsc{Clumpy}}
\newcommand{\thetabold}{\mbox{\boldmath$\theta$}}

\usepackage{graphicx}

\title[Constraining clumpy dusty torus models]{Constraining clumpy dusty torus models \\ using optimized filter sets}
\author[A. Asensio Ramos \& C. Ramos Almeida]
{\parbox{\textwidth}{A. Asensio Ramos$^{1,2}$, 
C. Ramos Almeida$^{1,2}$\thanks{E-mail: aasensio@iac.es, cra@iac.es}}\vspace{0.4cm}\\
\parbox{\textwidth}{$^{1}$Instituto de Astrof\'{\i}sica de Canarias, 
C/V\' ia L\'actea, s/n, E-38205, La Laguna, Tenerife, Spain\\
$^{2}$Departamento de Astrof\'{\i}sica, Universidad de La Laguna, E-38205 La Laguna, Tenerife, Spain}}

\begin{document}

\date{}

\pagerange{\pageref{firstpage}--\pageref{lastpage}} \pubyear{2012}

\maketitle

\label{firstpage}

\begin{abstract}
Recent success in explaining several properties of the dusty torus around the central engine
of active galactic nuclei has been gathered with the assumption of clumpiness. The properties
of such clumpy dusty tori can be inferred by analyzing spectral energy distributions (SEDs), sometimes with
scarce sampling given that large aperture telescopes and long integration times are
needed to get good spatial resolution and signal.
We aim at using the information already present in the data and the assumption of clumpy dusty
torus, in particular, the CLUMPY models of Nenkova et al., to evaluate the optimum next observation
such that we maximize the constraining power
of the new observed photometric point.
To this end, we use the existing and barely applied idea of Bayesian adaptive exploration,
a mixture of Bayesian inference,
prediction and decision theories. The result is that the new photometric filter to use is the 
one that maximizes the expected utility, which we approximate with the
entropy of the predictive distribution. In other words, we have to sample
where there is larger variability in the SEDs compatible with the data with what we know of the model
parameters.
We show that Bayesian adaptive exploration can be used to suggest new observations, and
ultimately optimal filter sets, 
to better constrain the parameters of the clumpy dusty torus models. In general, we find that the region 
between 10 and 200 $\mu$m produces the largest increase in the expected utility, although 
sub-mm data from ALMA also prove to be useful. 
It is important to note that here we are not considering the angular resolution of the data, which is
key when constraining torus parameters. Therefore, the expected utilities derived from this methodology 
must be weighted with the spatial resolution of the data.
\end{abstract}

\begin{keywords}
methods: data analysis, statistical --- galaxies: active, nuclei, Seyfert
\end{keywords}


\section{Introduction}

The unified model for active galaxies \citep{Antonucci93,Urry95} is based on the existence of a dusty toroidal structure 
surrounding the central region of active galactic nuclei (AGN). 
Considering this geometry of the obscuring material, the central engines of Type-1 AGN can be seen directly, 
resulting in typical spectra with both narrow and broad emission lines, whereas in Type-2 AGN the 
broad line region (BLR) is obscured.

The infrared (IR) range (and particularly the mid-infrared; MIR) is key to characterize the torus, 
since the dust reprocesses the optical and ultraviolet radiation generated in the accretion process
and re-emits it in this range. 
However, considering the small torus size\footnote{Less than 10 pc in the case of Seyfert galaxies based on 
ground-based MIR imaging (e.g. \citealt{Packham05,Radomski08} and interferometric observations 
(e.g. \citealt{Jaffe04,Tristram07}).}, high angular resolution turns to be essential to isolate as much
as possible its emission.

Pioneering work in modelling the dusty torus \citep{Pier92,Pier93} 
assumed a uniform dust density distribution to simplify the modelling, although from the start, \citet{Krolik88} 
realized that smooth dust distributions cannot survive within the immediate AGN vicinity.
To solve the various discrepancies between observations and models, 
an intensive search for an alternative torus geometry has been carried out in the last decade. 
The clumpy torus models \citep{Nenkova02,Nenkova08a,Nenkova08b,Honig06,Schartmann08}
propose that the dust is distributed in clumps, instead of homogeneously filling the torus volume. 
These models are making significant progress in accounting for the MIR emission of AGN 
\citep{Mason06,Mason09,Mor09,Horst08,Horst09,Nikutta09,Ramos09a,Ramos11a,Ramos11b,Honig10,Alonso11,Alonso12b,Alonso12a}.

In previous works we constructed subarcsecond resolution 
IR spectral energy distributions (SEDs) for about 20 nearby Seyfert galaxies and successfully reproduced them 
with the clumpy torus models of Nenkova et al. (hereafter CLUMPY). It is worth mentioning, however, that
some observational results show that the CLUMPY models alone cannot explain the IR SEDs of a sample of 
PG quasars \citep{Mor09,Mor12}. The latter authors needed an additional hot dust component to reproduce the SEDs. 
Moreover, recent interferometry results \citep{Kishimoto11,Honig12} indicate that a single component torus 
does not reproduce the observations.

The CLUMPY database now contains $\sim$5$\times$10$^6$ models, calculated for a fine grid of model
parameters. The inherent degeneracy between these parameters has to be taken into account
when fitting the observables. To this end, we developed the Bayesian inference tool \B.
Details on the interpolation methods and algorithms employed can be found in \citet{bayesclumpy09}. 
We point out that, given the specificities of the Bayesian inference approach
we follow, in the following analysis we will not be using the original set of models 
described in \citet{Nenkova08a,Nenkova08b}, but an interpolated version of them. 

In \citet{Ramos09a,Ramos11b} we fitted IR SEDs constructed using MIR nuclear 
fluxes obtained with 8 m telescopes and NIR measurements of similar resolution from the literature (see also
\citealt{Alonso11}). 
Some of the SEDs were well-sampled (e.g. the Circinus galaxy and Centaurus A) whilst others comprised 
just three photometric data points (e.g. NGC 1365 and NGC 1386). The better the sampling of the SED, 
the more constrained the torus parameters (see e.g. \citealt{Alonso12b}). Considering the need for 
8-10 m telescopes to isolate the torus emission, it is necessary to determine the minimum
number of filters required to constrain the model parameters. We utilize the output of our
code \B\ in a Bayesian experiment design framework. Our aim is to design
the experiment (observation of a source using a selected filter) that introduces more
constraints for the parameters of the CLUMPY models. Using our Bayesian approach, we
can investigate which and how many optical, IR, and sub-mm filters restrict the most the parameter space, as 
well as which wavelengths provide more information about each of the parameters. Although here we
present results for the models of Nenkova et al., the formalism can be applied
to any other set of models, including multi-component ones.
Thus, this work can be useful for the community when applying for telescope observations.

\section{Clumpy Dusty Torus Models and Bayesian approach}
\label{sec:models}
\begin{figure}
\centering
\includegraphics[width=8cm]{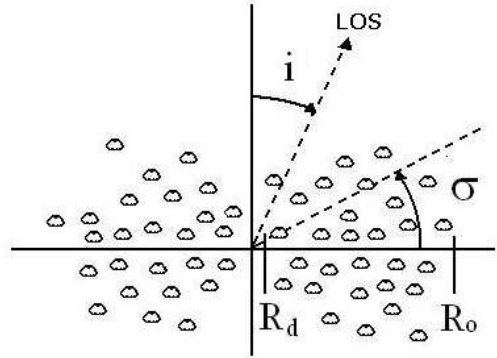}
\caption{Scheme of the clumpy torus described in \citet{Nenkova08a,Nenkova08b}. The radial extent of the torus is defined by the outer radius
($R_o$) and the dust sublimation radius ($R_d$). All the clouds are supposed to have the same $\tau_{V}$, and $\sigma$ 
characterizes the width of the angular distribution. The number of cloud encounters is function of the viewing angle, $i$.}
\label{clumpy_scheme}
\end{figure}

The CLUMPY models of \citet{Nenkova02} hold that the dust
surrounding the central engine of an AGN is distributed in clumps.  
These clumps are distributed with a radial extent $Y = R_{o}/R_{d}$, where 
$R_{o}$ and $R_{d}$ are the outer and inner radius of the toroidal distribution, respectively (see Figure \ref{clumpy_scheme}). 
The inner radius is defined by the dust sublimation temperature ($T_{sub} \approx 1500$ K),
with $R_{d} = 0.4~(1500~K~T_{sub}^{-1})^{2.6}(L / 10^{45}\,\mathrm{erg ~s^{-1}})^{0.5}$ pc.  
Within this geometry, each clump has the same optical depth ($\tau_{V}$).
The average number of clouds along a radial equatorial ray is $N_0$. The radial density profile is a
power-law ($\propto r^{-q}$). A width parameter, $\sigma$, characterizes the angular distribution of the clouds, which has
a smooth edge.  The number of clouds along the LOS 
at an inclination angle $i$ is $N_{LOS}(i) = N_0~e^{(-(i-90)^2/\sigma^2)}$. 
For a detailed description of the clumpy models see \citet{Nenkova02,Nenkova08a,Nenkova08b}.

In every Bayesian scheme, one needs to specify a-priori information about the model
parameters. This is done through the prior distribution \citep[see][for more details]{bayesclumpy09}. We consider them 
to be truncated uniform distributions for the six model parameters in the intervals reported in Table \ref{tab:parametros}. Therefore, we give the same weight
to all the values in each interval. 
To compare with the observations,
\B\ simulates the effect of the employed filters on the SED by integrating
the product of the synthetic SED and the filter transmission curve. Observational errors
are assumed to be Gaussian.

\begin{table}
\begin{minipage}{140mm}
\caption{Clumpy Model Parameters and Considered Intervals\label{tab:parametros}}
\begin{tabular}{lcl}
\hline
Parameter & Abbreviation & Interval \\
\hline
Width of the angular &         &      \\
\,\, distribution of clouds            	&  $\sigma$      &   [15\degr, 75\degr]   \\
Radial extent of the torus                             	& $Y$             & [5, 100]        	\\
Number of clouds along  		&            &  		\\
\,\, the radial equatorial direction 		&    $N_0$        &  [1, 15]		\\
Power-law index of the           	&             &     		\\
\,\,radial density profile          	&  $q$             &   [0, 3]  		\\
Inclination angle of the torus                         	& $i$             & [0\degr, 90\degr]      \\
Optical depth per single cloud                         	& $\tau_{V}$      & [5, 150] 		\\
\hline
\end{tabular}
\end{minipage}
\end{table}

The results of the fitting process of the IR SEDs with the 
CLUMPY models are the posterior distributions for the six free 
parameters that describe the models. 
When the observed data introduce sufficient information into the fit, 
the resulting probability distributions will clearly differ from the input uniform distributions, 
either showing trends or being centered at certain values within the intervals considered.

%
\section{Bayesian adaptive exploration}
\label{sec:bae}
In this section we briefly describe our approach to the experiment design, following the
lines of the theory of Bayesian adaptive exploration, as developed by \cite{sebastiani00} and \cite{loredo04}.
The output of this analysis is a measure of the ``utility'' of each one of the available
photometric filters to constrain the models introduced in Sec. \ref{sec:models}.

\subsection{General problem}
The problem at hand can be stated as follows. Let $\mathbf{D}$ be an $N$-dimensional vector containing observations
that sample an AGN SED using $N$ different filters. A fundamental
assumption that we have to make (inherent to any inference process) is to consider that the 
SEDs we are analyzing can be correctly reproduced with the CLUMPY models. 
If we have a set of $M$ new filters available (in our case, from the list of Table \ref{tab:filters}) to carry out new observations, we want to know 
which is the filter that, when we acquire a new observation with it, maximizes the amount of information that 
we gain about the model parameters. As stated before, this is a crucial problem to be solved
when applying for observing time in large-aperture telescopes, given the large overpetition of
such telescopes. In order to solve this problem, we apply a recent methodology of Bayesian experiment design 
termed Bayesian adaptive exploration (BAE), as presented by 
\cite{sebastiani00} and \cite{loredo04}. In essence, BAE can
be understood as an iteration of an \emph{observation}-\emph{inference}-\emph{design} scheme. Starting
from the current observations, we infer the model parameters using standard Bayesian techniques like
what can be achieved with \B\ \citep{bayesclumpy09}. The inference process
allows us to predict new data at the light of the information gained from the observations. From these
synthetic observations, it is possible to predict which new experiments will carry more information. The decision of
which is the best experiment to carry out is not an issue of Bayesian inference, but one has
to rely on the foundations of Bayesian decision theory \citep{berger85}. This procedure can
be iterated until convergence, although we will not pursue this issue here, only the suggestion of
a new observations. We summarize in the following the tools that we have used in this
paper.

\begin{table}
\tiny
\begin{minipage}{140mm}
\caption{Filters considered in this work\label{tab:filters}}
\begin{tabular}{ccccc}
\hline
ID & Instrument & Telescope & $\lambda_0$ [$\mu$m] & FWHM [$\mu$m] \\
\hline
F606W 	     & WFPC2	 & HST 		& 0.603 	 & 0.22	\\                
F791W 	     & WFPC2     & HST 		& 0.806 	 & 0.19	\\                
F110W 	     & NICMOS	 & HST 		& 1.102 	 & 0.59 \\                
F160W 	     & NICMOS	 & HST 		& 1.595 	 & 0.40	\\                
F187N        & NICMOS	 & HST 		& 1.874          & 0.02	\\                
F222M 	     & NICMOS    & HST 		& 2.216 	 & 0.14	\\                  
J	     & NACO      & VLT          & 1.265	 	 & 0.25	\\                  
H	     & NACO      & VLT          & 1.659	 	 & 0.34	\\             
Ks	     & NACO      & VLT          & 2.180 	 & 0.35	\\ 
IB2.42	     & NACO      & VLT          & 2.431		 & 0.06	\\             
L'	     & NACO      & VLT          & 3.805 	 & 0.63	\\             
IB4.05       & NACO      & VLT          & 4.056 	 & 0.06 \\             
M'	     & NACO      & VLT          & 4.781 	 & 0.60	\\             
Js	     & ISAAC     & VLT    	& 1.249	         & 0.16	\\             
H	     & ISAAC     & VLT    	& 1.652	         & 0.30 \\             
Ks	     & ISAAC     & VLT    	& 2.164	         & 0.27	\\             
L	     & ISAAC     & VLT    	& 3.779	         & 0.58	\\             
Nb\_M	     & ISAAC     & VLT    	& 4.657	         & 0.10	\\
Js	     & SOFI      & NTT 		& 1.249 	 & 0.16	\\             
H	     & SOFI      & NTT 		& 1.652 	 & 0.30	\\                   
Ks	     & SOFI      & NTT 		& 2.164 	 & 0.27	\\                   
J	     & IRCAM3    & UKIRT	& 1.248 	 & 0.16	\\                   
H	     & IRCAM3    & UKIRT	& 1.630 	 & 0.30	\\                   
K	     & IRCAM3    & UKIRT	& 2.200 	 & 0.34	\\                           
L'	     & IRCAM3    & UKIRT	& 3.774 	 & 0.70	\\                           
M	     & IRCAM3    & UKIRT	& 4.758 	 & 0.64	\\                           
H 	     & IRAC-1	 & 2.2m ESO     & 1.651 	 & 0.30	\\                       
K 	     & IRAC-1	 & 2.2m ESO     & 2.161 	 & 0.27	\\                       
K' 	     & NSFCam	 & IRTF		& 2.113 	 & 0.34	\\                                
L 	     & NSFCam	 & IRTF		& 3.498 	 & 0.61	\\                                
K            & TIMMI2    & 3.6m ESO     & 2.161 	 & 0.27	\\                  
L            & TIMMI2    & 3.6m ESO     & 3.498 	 & 0.58	\\                  
M            & TIMMI2    & 3.6m ESO     & 4.758 	 & 0.64	\\                  
Si2   	     & MICHELLE  & Gemini N & 9.150 	 & 0.87	\\               
N     	     & MICHELLE  & Gemini N & 10.50	 	 & 5.58	\\               
Si4   	     & MICHELLE	 & Gemini N & 10.54 	 & 1.01	\\                 
N'    	     & MICHELLE	 & Gemini N & 11.30 	 & 2.40	\\                 
Si6   	     & MICHELLE	 & Gemini N & 12.72 	 & 1.16	\\                 
Qa    	     & MICHELLE	 & Gemini N & 18.47 	 & 1.95	\\
Q    	     & MICHELLE	 & Gemini N & 20.86 	 & 8.97	\\
Si2	     & T-ReCS    & Gemini S & 8.725 	 & 0.78	\\        
N	     & T-ReCS    & Gemini S & 10.31 	 & 5.24	\\        
Np	     & T-ReCS    & Gemini S & 11.35 	 & 2.27 \\        
Si5          & T-ReCS    & Gemini S & 11.65          & 1.16 \\        
Qa	     & T-ReCS    & Gemini S & 18.34 	 & 1.52	\\
N	     & OSCIR     & CTIO         & 10.82 	 & 5.16	\\                  
IHW18	     & OSCIR     & CTIO         & 18.12 	 & 1.61	\\
ArIII  	     & VISIR	 & VLT          & 8.996 	 & 0.13	\\               
SIV  	     & VISIR	 & VLT          & 10.46 	 & 0.16	\\               
PAH2  	     & VISIR	 & VLT          & 11.27 	 & 0.59	\\               
PAH2\_2       & VISIR	 & VLT          & 11.74	 	 & 0.37	\\               
NeII\_1       & VISIR	 & VLT          & 12.27 	 & 0.19	\\               
NeII\_2       & VISIR	 & VLT          & 13.04 	 & 0.22	\\               
Q2  	     & VISIR	 & VLT          & 18.75 	 & 0.86	\\ 
CC\_Si2       & CanariCam  & GTC         & 8.67          & 1.04 \\
CC\_N        & CanariCam  & GTC          & 10.31          & 5.25 \\
CC\_Q4       & CanariCam  & GTC          & 20.39          & 0.97 \\
CC\_Q8       & CanariCam  & GTC          & 24.53          & 0.75 \\
LWC$_-$31.5  & FORCAST   & SOFIA     & 31.38		 & 4.54      \\ 
LWC$_-$33.6  & FORCAST   & SOFIA     & 33.60		 & 1.56 	 \\
LWC$_-$34.8  & FORCAST   & SOFIA     & 34.81		 & 3.42 	 \\
LWC$_-$37.1  & FORCAST   & SOFIA     & 37.18		 & 2.13 	 \\
PACS70 	     & PACS      & Herschel   & 71.07 	 & 20.9 \\          
PACS100      & PACS	 & Herschel   & 102.2 	 & 36.0 \\          
PACS160      & PACS	 & Herschel   & 165.9 	 & 74.5 \\          
SPIRE250     & SPIRE     & Herschel   & 251.6 	 & 77.9 \\          
SPIRE350     & SPIRE     & Herschel   & 353.3 	 & 106.6 \\          
SPIRE500     & SPIRE     & Herschel   & 508.8 	 & 198.2 \\
ALMA Ch9     & \dots       & ALMA         & 460.0 	 & 77.3 \\
ALMA Ch7     & \dots       & ALMA         & 945.0 	 & 280.2 \\
\hline
\end{tabular}
\end{minipage}
\end{table}

Let $\thetabold$ be the vector with the parameters that define the CLUMPY model. Once the set of
observations $\mathbf{D}$ for the current filters have been acquired, the role of Bayesian inference is to compute
the posterior distribution $p(\thetabold|\mathbf{D},I)$ that encodes all the information
we possess about the model parameters at the light of the observations and assuming our
current experiment $I$. Using the posterior distribution, it is possible to predict which
is the expected value of a fictitiuos new observation $o$ at filter $f$ under
the experiment $I_f$ using the well-known
predictive distribution \citep[e.g.,][]{bishop06}:
\begin{eqnarray}
p(o|\mathbf{D},I_f) &=& \int d\thetabold \, p(o,\thetabold|\mathbf{D},I_f) \nonumber \\
&=& \int d\thetabold \, p(o|\thetabold,\mathbf{D},I_f) \, p(\thetabold|\mathbf{D},I_f),
\label{eq:predictive_distribution}
\end{eqnarray}
where the conditioning on $\mathbf{D}$ of $p(o|\thetabold,\mathbf{D},I_f)$ is, in fact, irrelevant once $\thetabold$ is known.
The term $p(o|\thetabold,\mathbf{D},I_f)$ is just the likelihood of getting the new
observation $o$.
According to the Bayesian decision theory, one should choose the filter $f$ that maximizes the
\emph{expected utility} (EU), defined as:
\begin{equation}
EU(f) = \int do \, p(o|\mathbf{D},I_f) \, U(o,f),
\label{eq:expected_utility}
\end{equation}
where the function $U(o,f)$ is the \emph{utility}, a function that is at the core of decision
theory and that quantifies the information gain of a new experiment and can also include
the potential cost of the new experiment at filter $f$ (i.e., we could potentially include factors
that, e.g., increase the weight of filters that provide a better spatial resolution or that
takes into account the difficulty of being awarded with observing with a certain telescope).
In other words, we have to choose the filter that maximizes the
value of the expected utility using the predictions of the flux at filter $f$ according to our
current knowledge of the model.

\begin{table*}
\begin{minipage}{140mm}
\caption{Measured flux densities\label{tab:fluxes}}
\begin{tabular}{l|cccccc}
\hline
Filters & NGC 1566 & NGC 4151 & Circinus & NGC 1068 & avg Sy1 & avg Sy2 \\
\hline
NICMOS F110W &    $-$         &  60$\pm$6      &     $-$           &  9.8$\pm$2      &       $-$           &  $-$  \\
NICMOS F160W &    $-$         &  100$\pm$10    &  1.6$\pm$0.2   & 98$\pm$15       &  0.05$\pm$0.01   &  0.001$\pm$0.001\\
NICMOS F222M &    $-$         &  197$\pm$20    &     $-$           & 445$\pm$100     &       $-$           & $-$\\
NACO J       & 1.1$\pm$0.1 &      $-$          &  4.8$\pm$0.7   &     $-$            &  0.07$\pm$0.03   &  0.003$\pm$0.002 \\
NACO Ks      & 2.1$\pm$0.1 &      $-$          &   19$\pm$2     &     $-$            &  0.12$\pm$0.05   &  0.023$\pm$0.018 \\
NACO 2.42    &             &      $-$          &  31$\pm$3      &     $-$            &        $-$          & $-$\\
NACO L'      & 7.8$\pm$0.1 &      $-$          &  380$\pm$38    &     $-$            &  0.35$\pm$0.11   &  0.15$\pm$0.09 \\
NACO M'      &             &      $-$          &  1900$\pm$190  &     $-$            &  0.48$\pm$0.13   &  0.31$\pm$0.09 \\
NSFCam L     &    $-$         &  325$\pm$65    &      $-$          &      $-$           &       $-$           &$-$\\
IRCAM3 M     &    $-$         &  449$\pm$34    &      $-$          &  2270$\pm$341   &         $-$         &$-$\\
TIMMI2 L     &    $-$         &      $-$          &    $-$            & 920$\pm$138     &       $-$           & $-$ \\
OSCIR N      &    $-$         & 1320$\pm$200   &      $-$          &      $-$           &      $-$            &$-$\\
OSCIR IHW18  &    $-$         & 3200$\pm$800   &      $-$          &      $-$           &      $-$            &$-$\\
T-ReCS Si2   & 29$\pm$4    &      $-$          & 5620$\pm$843   & 10000$\pm$1500  &  1.00$\pm$0.15   &  1.00$\pm$0.15\\ 
T-ReCS Qa    & 63$\pm$9    &      $-$          & 12791$\pm$3198 & 21773$\pm$5443  &  4.46$\pm$2.19   &  4.36$\pm$1.93\\
VISIR PAH2\_2& 117$\pm$29  &      $-$          &      $-$          &      $-$           &       $-$           &$-$\\
VISIR Q2     & 128$\pm$32  &      $-$          &      $-$          &      $-$           &       $-$           &$-$\\
\hline
\end{tabular}
\medskip
Fluxes in mJy from \cite{Ramos09a,Ramos11b} and \cite{Alonso11}. In the case of the Sy1 and Sy2 templates, 
the fluxes have been normalized to the central wavelength of the T-ReCS Si2 filter.
\end{minipage}
\end{table*}

The previous definitions are somewhat obvious and the core of Bayesian decision theory is 
located on the appropriate definition of the utility function.
We follow here the suggestion of \cite{lindley56} \citep[see also][]{loredo04}, who
suggested that, since we want to maximize the information about the model parameters $\thetabold$,
it makes sense to use the information encoded on the posterior for $\thetabold$, as measured
by the negative differential entropy:
\begin{equation}
U(o,f) = -H[\thetabold|o,\mathbf{D},I_f],
\label{eq:utility}
\end{equation}
where the notation means that the entropy is computed for the posterior for the distribution $p(\thetabold|o,\mathbf{D},I_f)$.
The differential entropy is, following the standard definition, given by:
\begin{equation}
H[\thetabold|o,\mathbf{D},I_f] = -\int d\thetabold \, p(\thetabold|o,\mathbf{D},I_f) \log p(\thetabold|o,\mathbf{D},I_f).
\label{eq:entropy_def}
\end{equation}
Consequently, we have to compute the following quantity for all the available filters
\begin{equation}
\mathrm{EU}(f) = \int \int do \, d\thetabold \, p(o|\mathbf{D},I_f) \,p(\thetabold|o,\mathbf{D},I_f) \log p(\thetabold|o,\mathbf{D},I_f),
\end{equation}
and choose the filter $f$ that maximizes it. One could plug the expression for the predictive
distribution of Eq. (\ref{eq:predictive_distribution}) on the expected utility and end
up with a triple multidimensional integral that needs to be computed for each value of $f$. 
However, it turns out to be much easier to plug Eqs. (\ref{eq:utility}) and (\ref{eq:entropy_def}) onto Eq. (\ref{eq:expected_utility}) and
apply Bayes theorem \citep[e.g.,][]{gregory05} to $\log p(\thetabold|o,\mathbf{D},I_f)$, so that we have to end up with:
\begin{eqnarray}
\mathrm{EU}(f) &=& \int do \, d\thetabold \, p(\thetabold|\mathbf{D},I_f) \,p(o|\thetabold,\mathbf{D},I_f) \log p(o|\thetabold,\mathbf{D},I_f) \nonumber \\
&+& \int do \, d\thetabold \, p(o|\thetabold,\mathbf{D},I_f) \, p(\thetabold|\mathbf{D},I_f) \log p(\thetabold|\mathbf{D},I_f) \nonumber \\
&-& \int do \, d\thetabold \, p(\thetabold|\mathbf{D},I_f) \, p(o|\mathbf{D},I_f) \log p(o|\mathbf{D},I_f).
\end{eqnarray}
Using the definition of entropy of Eq. (\ref{eq:entropy_def}), we can rewrite the previous expression as:
\begin{eqnarray}
\mathrm{EU}(f) &=& -\int d\thetabold \, p(\thetabold|\mathbf{D},I_f) H[o|\thetabold,\mathbf{D},I_f] \nonumber \\
&-& \int do \, p(o|\thetabold,\mathbf{D},I_f) H[\thetabold|\mathbf{D},I_f] \nonumber \\
&+& \int d\thetabold \, p(\thetabold|\mathbf{D},I_f) H[o|\mathbf{D},I_f].
\end{eqnarray}
Note that the entropies $H[\thetabold|\mathbf{D},I_f]$ and $H[o|\mathbf{D},I_f]$ can be extracted
from the integrals because they do not depend on the integration variable. Given that the probability distributions are normalized to unit area, the expression
simplifies to:
\begin{eqnarray}
\mathrm{EU}(f) &=& -\int d\thetabold p(\thetabold|\mathbf{D},I_f) H[o|\thetabold,\mathbf{D},I_f] \nonumber \\
&-& H[\thetabold|\mathbf{D},I_f] + H[o|\mathbf{D},I_f].
\label{eq:eu}
\end{eqnarray}

\subsection{Simplifications}
Clearly, the second term (the entropy of the posterior distribution considering the existing
data) is independent of the election of the new filter, so it becomes constant with respect to $f$ and
can be dropped from the computation.                                                                                                   
The term $H[o|\thetabold,\mathbf{D},I_f]$, which is the entropy of the likelihood for the new observed point
at filter $f$, can be assumed to be constant if the expected noise is independent of the
measurement. This is the case when the measurement for the new filter $f$ is done under the presence
of additive (Gaussian) noise with a variance that is independent of $f$. However, this is not
usually the case since the noise variance at different wavelengths can vary a lot. Assuming that 
the observation $o$ is perturbed with Gaussian noise with variance $\sigma_f$, the likelihood 
$p(o|\thetabold,\mathbf{D},I_f)$ is a Gaussian distribution, so its entropy is analytically expressed as:
\begin{equation}
H[o|\thetabold,\mathbf{D},I_f] = \frac{1}{2} \log \left( 2\pi \mathrm{e} \sigma_f^2 \right).
\end{equation}
Given that the previous entropy is independent of $\thetabold$, it can be extracted from the 
integral in Eq. (\ref{eq:eu}) and use the fact that the posterior is normalized to unit area. Therefore, 
the final expression for the expected utility that we use in this work simplifies to:
\begin{equation}
\mathrm{EU}(f) = -\frac{1}{2} \log \left( 2\pi \mathrm{e} \sigma_f^2 \right) - \int do \, p(o|\mathbf{D},I_f) \log p(o|\mathbf{D},I_f),
\label{eq:eu_final}
\end{equation}
with $p(o|\mathbf{D},I_f)$ given by Eq. (\ref{eq:predictive_distribution}). Note that if the noise variance 
does not depend on $f$, we end up with the expected utility
derived by \cite{sebastiani00} and \cite{loredo04}. They noted that the previous expression leads to 
a maximum entropy sampling, so that we select the new filter where we know
the least. In other words, we should sample where the current values of the model parameters allow for
a large variability of the SED.

\begin{figure*}
\centering
\includegraphics[width=0.75\textwidth]{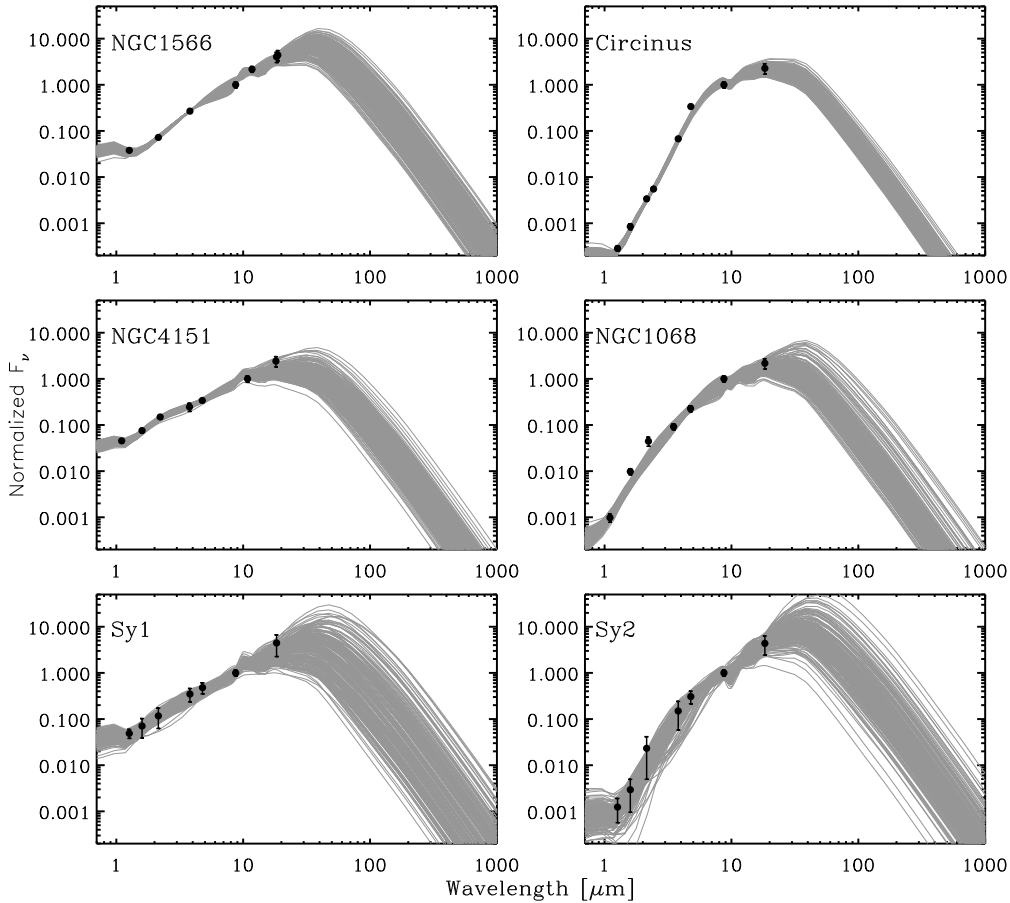}
\caption{IR SEDs used in this paper. The circles with the error bars are the 
observed points and they are drawn at the central wavelength of each filter. The left panels correspond
to Sy1 galaxies, while the right panels are Sy2. The average Sy1 and Sy2 SEDs have been obtained by a
simple mean of galaxies belonging to each group once their SEDs have been interpolated onto the filters in which
Circinus was observed (REF). Fluxes are normalized to the data point closer to 10 $\mu$m. The grey
curves are the CLUMPY SEDs that have been sampled from the posterior.}
\label{fig:seds}
\end{figure*}

\subsection{Technicalities}
\label{sec:technical}
The computation of the integral of Eq. (\ref{eq:eu_final}) can be done following \cite{loredo04},
who used a Monte Carlo estimation technique. First, we can compute a Monte Carlo estimation of $p(o|\mathbf{D},I_f)$ using:
\begin{equation}
p(o|\mathbf{D},I_f) \approx \frac{1}{N_\mathrm{pos}} \sum_{j=1}^{N_\mathrm{pos}} p(o|\thetabold_j,\mathbf{D},I_f),
\label{eq:montecarlo_predictive}
\end{equation}
where the $\thetabold_j$ are $N_\mathrm{pos}$ samples from the posterior distribution $p(\thetabold|\mathbf{D},I_f)$
that have been previously computed with \B\ for the current set of observations. Then, the
integral of Eq. (\ref{eq:eu_final}), the entropy of the predictive distribution, is computed using another Monte Carlo estimation:
\begin{equation}
H[o|\mathbf{D},I_f] \approx \frac{1}{N_\mathrm{pred}} \sum_{j=1}^{N_\mathrm{pred}} \log p(o_j|\mathbf{D},I_f),
\end{equation}
where the $o_j$ are samples from $p(o|\mathbf{D},I_f)$. However, we have tested that better results are
obtained if a simple trapezoidal quadrature is used to compute the integral of 
Eq. (\ref{eq:eu_final}). To this end, we discretize the
posible range of variation of $o$ in bins. For each bin, we compute $p(o|\mathbf{D},I_f)$
using the Monte Carlo estimation of Eq. (\ref{eq:montecarlo_predictive}). At the end, we compute the
entropy of the predictive distribution as:
\begin{equation}
H[o|\mathbf{D},I_f] \approx \sum_j w_j p(o_j|\mathbf{D},I_f) \log p(o_j|\mathbf{D},I_f),
\end{equation}
with $w_j$ the standard trapezoidal weights \citep[e.g.,][]{numerical_recipes86}.

\section{Results}
The Bayesian Adaptive Exploration theory, as summarized in the previous section, is applied
to the case of SEDs generated by the emission of a clumpy torus
around an AGN. Under the assumption that the parametric models of \cite{Nenkova08a,Nenkova08b} are representative of the
underlying physics, our analysis allows us to search for the filter that introduces more restrictions 
on the model parameters. 

Our set of potential filters is listed in Table \ref{tab:filters}. Our aim is to consider a sufficiently
complete list of photometric filters available in many medium- to large-aperture telescopes. In particular, 
we are considering all the optical, NIR, MIR and far-IR (FIR) filters that we have employed in our previous work, plus 
four new MIR filters from the CanariCam instrument, which has recently started to operate
at the 10 m Gran Telescopio Canarias (GTC), and another four filters from the Long Wavelength Camera (LWC) on the
FORCAST instrument. FORCAST is a MIR/FIR camera for the Stratospheric Observatory For Infrared Astronomy (SOFIA;
\citealt{Young12}). Finally, to inspect whether or 
not sub-mm data can set any constraints on the model parameters, we have also
added channels 7 and 9 of ALMA, although they cannot be considered filters per se. They are simulated 
as a square transmission efficiency with a peak
of 80\% in the ranges $[800,1090]$ $\mu$m for channel 7 and $[420,500]$ $\mu$m for channel 9.
Table \ref{tab:filters}  
presents the filter identification, together with the instrument and telescope in which they
are mounted. Additionally, we have tabulated the central wavelength of the filter, computed as
$\lambda_0=\int \lambda R(\lambda) d\lambda / \int R(\lambda) d\lambda$, where $R(\lambda)$ is the
transmission efficiency of the filter, and the full-width at half maximum (FWHM). The transmission
efficiency of the filters have been obtained from the literature. 

It is important to note that, for the sake of simplicity, in this work we are not taking into account  
the angular resolution provided by the different instruments considered here. In other words, we 
are assuming that all instruments listed in Table \ref{tab:filters} are equivalent except in the wavelength 
coverage. However, the reader must keep in mind that high spatial resolution is mandatory when 
trying to isolate the torus emission. Thus, aside from the expected utilities derived here for the different
instruments, it is important to consider the resolution when deciding which should 
be the next observation.

\begin{figure*}
\centering
\includegraphics[width=0.8\textwidth]{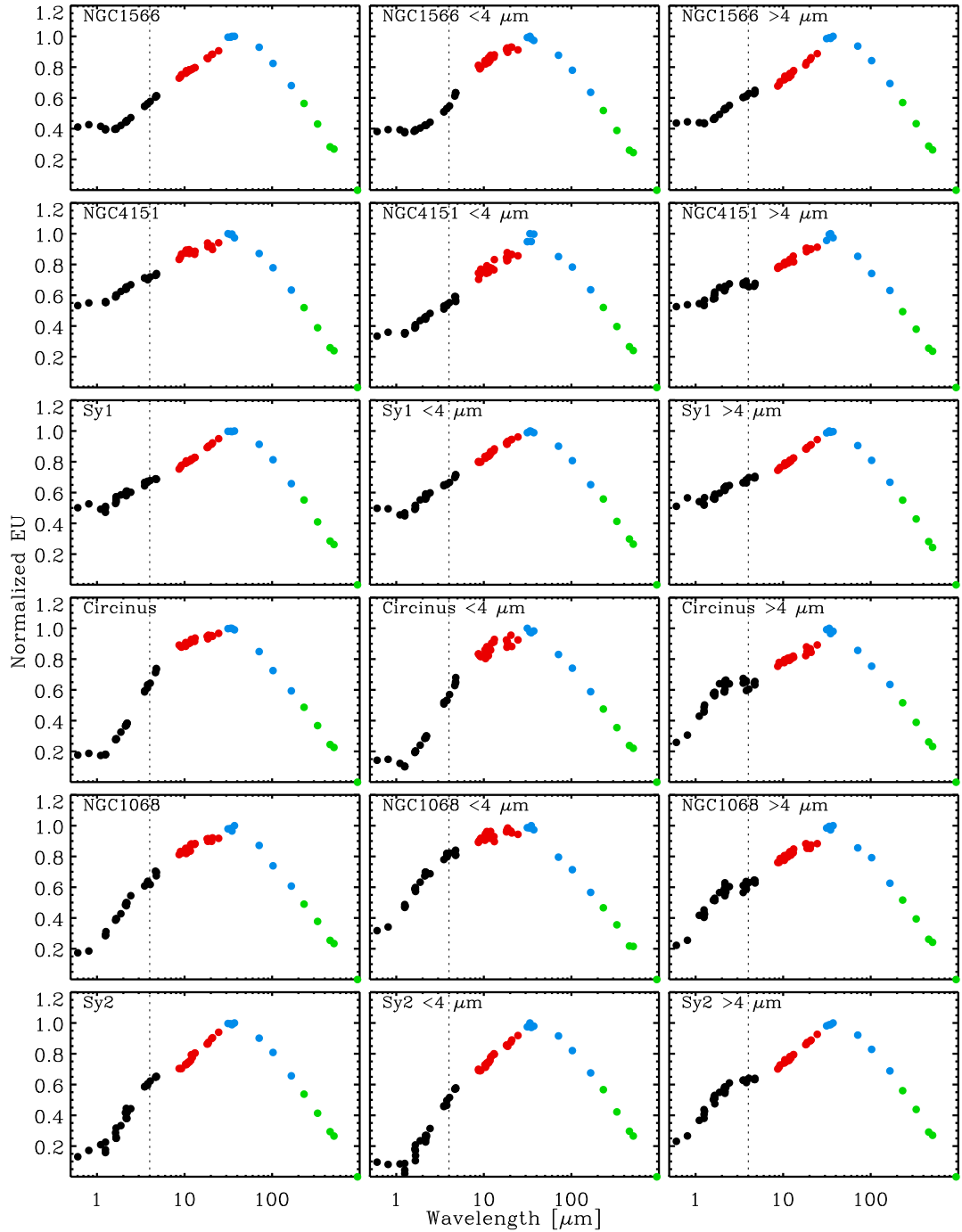}
\caption{Expected utility (EU) computed using Eq. (\ref{eq:eu}) and normalized to the maximum of each panel. EU
is calculated for all filters not present in the observations. The first column refers to the case in which 
we use all the available observations of the galaxy. The central and right columns refer to the cases
in which we only keep observations below and above 4 $\mu$m, respectively. For an easier visualization,
the different wavelength ranges have been color-coded.} Black dots correspond to 
filters with $\lambda_0 \leq 6$ $\mu$m, red dots to $6 < \lambda_0 \leq 25$ $\mu$m, blue dots to
$25 < \lambda_0 \leq 200$ $\mu$m and green dots to $\lambda_0 > 200$ $\mu$m.
\label{fig:expected_utility}
\end{figure*}

To demonstrate our approach, we have used four SEDs, two representative of Seyfert 1 (Sy1) and
another two representative of Seyfert 2 (Sy2) galaxies. These SEDs have been discussed by 
\cite{Ramos09a,Ramos11b} and \cite{Alonso11} using a 
deep Bayesian analysis. Additionally, here we use the Sy1 and Sy2 average templates presented 
in \cite{Ramos11b}. These mean templates were constructed using individual Sy1 and Sy2 SEDs 
of angular resolution $\la$0.55 arcsec, that were first interpolated onto the Circinus wavelength 
grid (1.265, 1.60, 2.18, 3.80, 4.80, 8.74, and 18.3 $\mu$m), 
and then normalized to the central wavelength of the T-ReCS Si2 filter (8.74 $\mu$m). 
These flux densities and
their associated errors are shown in Table \ref{tab:fluxes} and are displayed in graphical
form in Fig. \ref{fig:seds}, normalized to the filter that is closer to 10 $\mu$m. Note that the SEDs that we have selected
have quite dense samplings, as opposed to other sparsely sampled ones \citep{Ramos09a,Ramos11b}. 

\begin{figure*}
\centering
\includegraphics[width=0.49\textwidth]{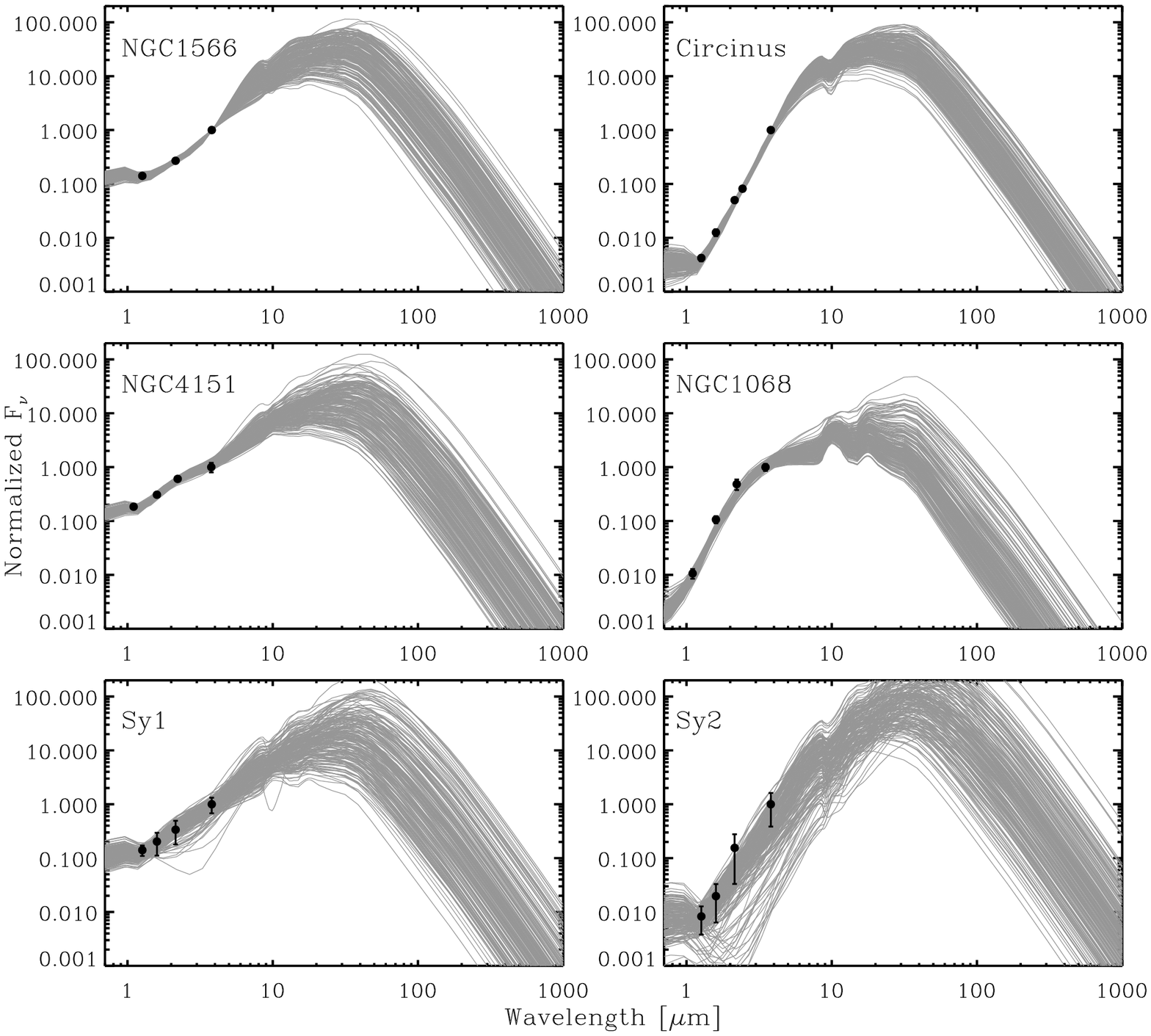}\includegraphics[width=0.49\textwidth]{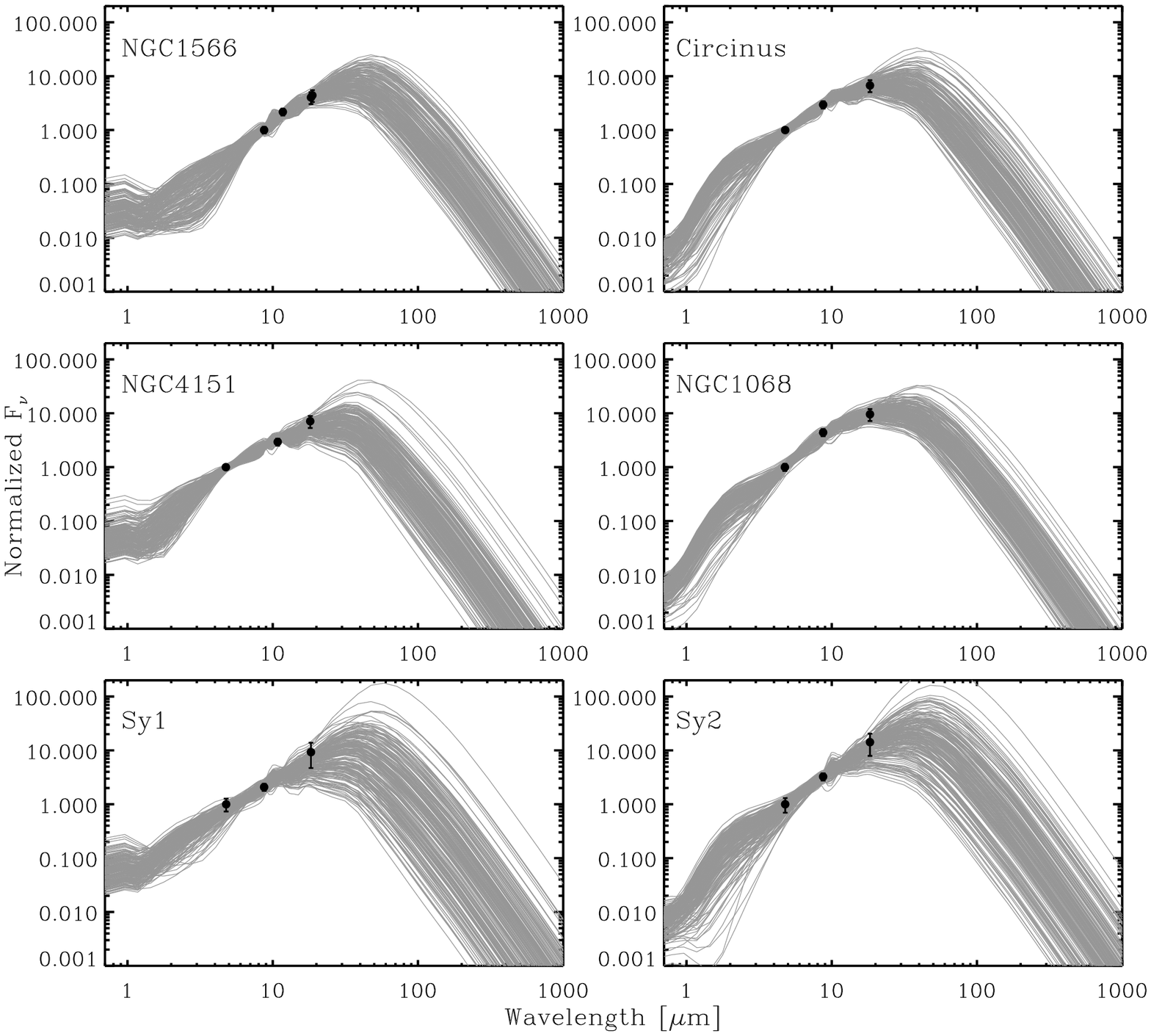}
\caption{Same as in Fig. \ref{fig:seds}, but considering observed data point below and above 4 $\mu m$ from the 
complete SEDs (6 left and 6 right panels respectively).}
\label{fig:seds_above_below}
\end{figure*}

For each SED we have carried out a full Bayesian analysis using \B. This analysis relies on sampling the
posterior distribution function as a function of the model parameters (shown in Table \ref{tab:parametros}).
We have used flat priors for all model parameters. For the Sy1 galaxies, we have additionally
included an AGN spectrum, which is added to the CLUMPY SED. The shape
of the AGN spectrum is the same that is considered for the infrared pumping of the dust in the torus \citep{Nenkova08a}.
Likewise, we also take into account a certain amount of extinction following the law of \cite{chiar_tielens00}. This
introduces the free parameter $A_V$, which is included into the inference for Sy1 galaxies with a flat prior in the
range $[0,5]$ mag. The inclusion of these two ingredients turns out to be essential for obtaining reliable
results for Sy1 sources.

\B\  generates samples of model parameters distributed according to the posterior distribution. Figure
\ref{fig:seds} shows, apart from the observed points, a representation of the SEDs reconstructed from the model parameters
sampled from the posterior. Note that the sampled SEDs have a small variability close to the observed
points with the smaller error bars. In principle, and according to the maximum entropy sampling
derived by \cite{sebastiani00} and \cite{loredo04}, the filter that we should select has to be located close
to the wavelength where the SEDs contain larger variability with the observational information we have acquired.
An evaluation of Eq.~(\ref{eq:eu_final}), following the ideas of Sec. \ref{sec:technical}, gives the results displayed
in the left column of Fig. \ref{fig:expected_utility}. This figure represents the value of $\mathrm{EU}(f)$ for filter
$f$ versus the central wavelength of the filter. The noise associated to any new observation, $\sigma_f$, depends 
on the wavelength: we consider 10\% of the value of the flux at $\lambda\leq 5~\mu m$, 20\% at $5~\mu m<\lambda<25~\mu m$
and 30\% at $\lambda\geq 25~\mu m$. These are the typical uncertainties that we measured for individual galaxies
in our previous work \citep[e.g.,][]{Ramos09a,Ramos11a,Ramos11b,Alonso11}.

The expected utility
is usually measured in terms of ``nats'' (or bits), i.e., units of information. However, since we are not interested
in their actual value but in relative measures, we normalize the plots to the maximum. Several properties are 
evident. First, the lack of a vertical
spread on the plots for a given wavelength means that the expected utility of the filters depends
essentially on the wavelength. Using different filters with the same central wavelength but different 
transmission properties produce the same increase in the information acquired. In other words, the
results are not dependent on the exact shape of the filter transmission curve. Obviously, other
reasons (atmospheric transmission, telescope diameter, availability, etc.) would make one particular 
filter preferable among a set of equivalent filters. However, under the same conditions, they introduce 
roughly the same information for constraining CLUMPY models. 

In general, the region between 10 and 200 $\mu$m produces
the largest increase in the expected utility. 
Out of the filters considered here, those from SOFIA provide the largest constraining power
for the CLUMPY models. However, the poor spatial resolution of SOFIA ($\sim$3--4 arcsec) represents 
a clear drawback and it might be preferable to choose other filters
with slightly smaller expected utility but better spatial resolution. 
The filters in the Q-band, especially the CanariCam Q8 filter 
($\lambda_0=$24.5 $\mu$m), are the next most useful ones for constraining 
the clumpy torus model parameters, specially if we consider their good spatial resolution 
($\sim$0.55 arcsec). They are closely followed by filters in the N-band ($\sim$8-12 $\mu$m)
and by the Herschel Space Observatory PACS filters, especially PACS70 and PACS100. 
The previous results are not surprising 
considering that the bulk of torus emission is observed in the MIR, peaking
at $\sim$20 $\mu$m. The PACS observations, which cover the spectral range from 
70 to 160 $\mu$m, are sampling cooler dust
within the torus, and thus helping to constrain parameters such as the torus radial extent ($Y$) 
and the radial distribution of the clouds ($q$), as discussed in \cite{Ramos11a}.

According to our results, data points obtained with the SPIRE instrument on Herschel 
--despite its low spatial resolution, which we are neglecting here-- 
still have a relatively large expected utility, and the same happens with sub-mm data from ALMA. We have 
considered two ALMA channels (7 and 9)
and the expected utility of channel 9 ($\lambda_0=460$ $\mu$m) is comparable to that provided 
by the NIR data in the case of the Sy2 galaxies. This 
result is encouraging for potential ALMA users looking for constraints on torus 
properties. ALMA will observe in the sub-mm regime with unprecedented 
angular resolution, and according to the analysis presented here, data obtained with 
channel 9 can provide useful constraints on torus model parameters. The case of Herschel 
SPIRE and ALMA clearly illustrates what we discussed at the beginning of this section:
although the expected utilities measured for the SPIRE filters are slightly higher than those
for ALMA channel 9, the angular resolution provided by ALMA makes it a better choice 
than SPIRE to sample the cool dust within the torus. On the other hand, the expected utility of channel 7 
($\lambda_0=945$ $\mu$m) is negligible in all the cases considered.

It is interesting to note that there is no apparent difference between Sy1 and Sy2 galaxies regarding 
the proposed new filters at wavelengths $\ga$4 $\mu$m. Therefore, it seems that spectral 
features that are supposed to serve to constrain the
model parameters because they are generally different in Sy1 and Sy2 (such as the silicate feature 
at 10 $\mu$m, generally in emission in Sy1 and in absorption in Sy2) do not make any difference in
suggesting new observations. It is worth 
noting, though, that here we are using photometry only, which can provide just an insight of spectral 
features such as the silicate band. The use of high angular resolution spectroscopy in this spectral region 
(provided by ground-based instruments such as VISIR, MICHELLE and CanariCam) can set constraints on 
the clumpy torus models that the photometry alone cannot \citep{Alonso11}. Thus, the use of spectroscopy 
would possibly change the suggested observations. Unfortunately, the expression for the expected
utility of Eq. (\ref{eq:eu_final}) in the spectroscopic case requires the computation of a multidimensional 
integral because the expected observation $o$ transforms into the vector $\mathbf{o}$. Tailor-made algorithms 
which are computationally heavy have to be applied to compute this integral. For this reason, we defer 
the spectroscopic case to a later study.

\begin{figure*}
\centering
\includegraphics[width=0.8\textwidth]{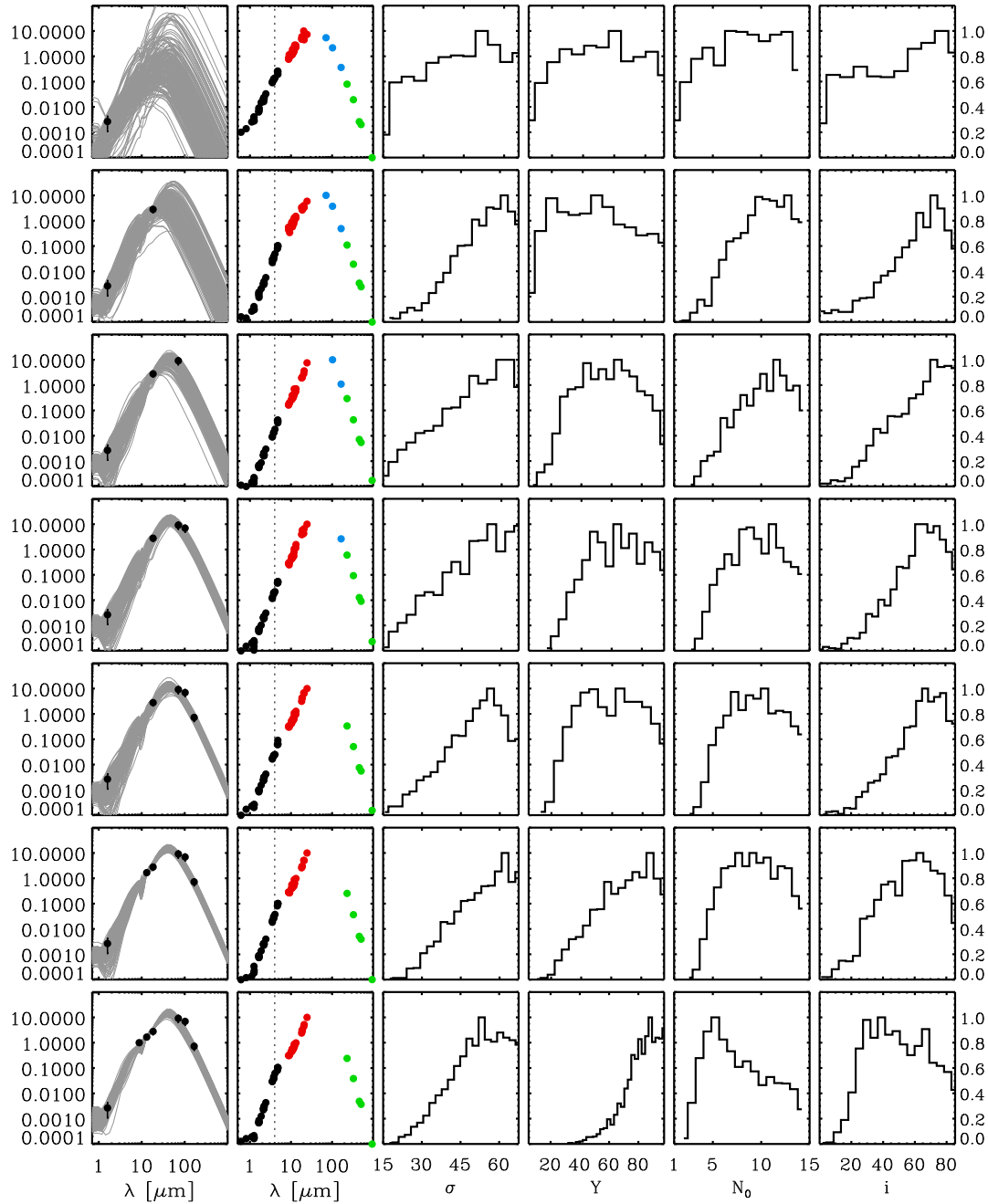}
\caption{Simulation of a Bayesian adaptive exploration experiment. A new observed point is added in each
column from those available for NGC 3081 (\protect\citealt{Ramos11a}). The first column shows the observations with
their associated error bars, together with the SEDs obtained from the posterior distribution in each case. The second
column shows the expected utility displaying the next best filter. The next four columns represent the
marginal posteriors of the CLUMPY parameters $\sigma$, $Y$, $N_0$ and $i$, respectively.}
\label{fig:experiment}
\end{figure*}

At wavelengths $\la$4 $\mu$m we start to see a difference between the expected utilities of 
Sy1 and Sy2. Optical and NIR observations help constraining the torus parameters for Sy1 (expected utilities between 
0.4 and 0.8) more than for Sy2 (expected utilities $\sim$0.2 around 1 $\mu$m). The hot dust emission from the directly 
illuminated faces of the clumps close to the central engine and the direct AGN emission strongly flatten 
the Sy1 IR SEDs (see Figure \ref{fig:seds}). Thus, in the case of Sy1, more observations at short wavelengths
are required to constrain the shape of the SED in this region, that in turn constrains parameters such 
as the inclination angle of the torus ($i$).

As an experiment, we have simulated which is the influence of having a reduced set of observations. To this
aim, we have considered the observed points below and above 4 $\mu$m from the complete SEDs. This separation
was chosen to roughly separate the NIR and the MIR keeping a similar number of points in both spectral domains. 
The resulting SEDs are
shown in the left and right six panels of Fig. \ref{fig:seds_above_below} respectively, using the same 
normalization as in Fig. \ref{fig:seds}. Again, the grey curves are samples of SEDs using
the model parameters from the Bayesian inference. Despite the 
sparser sampling of the SEDs in the two cases, the results remain essentially similar (see central and right panels
of Fig \ref{fig:expected_utility}), 
with the expected utility peaking in the 10-200 $\mu$m region. The only difference in the results 
is a small increase in the expected utitility of the optical and NIR data when only $\lambda>$4 $\mu m$
data is considered in the Sy2 SEDs (bottom right panels in Figure \ref{fig:expected_utility}).

\section{Simulated iterated BAE for NGC 3081}

The process we have described until this point is obviously not complete. As already described in
Sec. \ref{sec:bae}, the Bayesian adaptive exploration scheme starts from observations, and then applies
Bayesian inference tools to design a new potentially informative observation. Since the typical timescale
of this process can be very large given the necessity to apply for observing time in large-aperture
telescopes, we have carried out a simulated full process in Fig. \ref{fig:experiment}. We have
used the SED of the Sy2 galaxy NGC 3081 presented in \cite{Ramos11a}. The reason for choosing this 
galaxy is the availability of Herschel PACS nuclear fluxes. As a first step, we have assumed 
that we only have the observation
in the NICMOS F160W filter and let the Bayesian adaptive exploration scheme select the
following filters. \B\  then uses this point to sample the posterior distribution and
obtain marginal posteriors for the clumpy model parameters. The observed point, together with the 
SEDs sampled from the posterior are shown in the first column of Fig. \ref{fig:experiment}. The second 
column shows the expected utility, while the remaining panels display the
marginal posterior for $\sigma$, $Y$, $N_0$ and $i$, respectively. 

In order to simulate
the next experiment, we have selected the next observation from the set of observed points of the
full SED that has the largest expected utility, which turns out to be the T-ReCS Qa data point
($\lambda_0=18.3$ $\mu$m). The full process is then repeated until having all the
observed points of the SED in the lowest panel. The order in which the remaining filters have
been selected is: PACS70, PACS100, PACS160, VISIR NeII$_-$2 ($\lambda_0=13.04$ $\mu$m) and finally, 
T-ReCS Si2 ($\lambda_0=8.72$ $\mu$m). Given that the differences in expected utility are quite
small and can change a little because of the inherently random character of the Montecarlo sampling
done with \B\, the specific order might change from run to run. In fact, this means that one could
choose a filter giving a large expected utility even though it is not the one giving the largest
value if observing with this filter is easier for some specific reason.

It is important to note how the marginal posteriors
are constrained when augmenting the number of observed points. Once there are two points in the NIR and 
MIR, the marginal posterior of $\sigma$ is very similar to what
we find when considering the full SED. This is not the case for $Y$ and $N_0$, that 
considerably change after introducing the FIR data and the N-band data points (NeII$_{-}$2 and Si2 filters). 
Finally, the FIR does not seem to make any difference to the inclination angle of the torus, $i$, but
the addition of the N-band data points --especially the Si2 filter-- result in a different posterior 
shape, favouring intermediate torus inclinations.


\section{Conclusions}
The CLUMPY models have recently gained much success on explaining the observed
IR SED of the inner parsecs of nearby Seyfert galaxies. Given the observational
difficulty (one needs to obtain good signal-to-noise IR photometry of sources at subarcsecond resolution, 
and ideally, spectroscopy), the sampling of the SEDs is generally quite scarce. Assuming that the 
CLUMPY models are successful in explaining the observed SED, we have applied a Bayesian
adaptive exploration scheme to propose new observations given the presently available information
about a source. The results quantitatively indicate that the region between 10 and 200 $\mu$m is
crucial, almost independently of the presently available
observed points. 
We also find that optical and NIR data have higher expected utilities in the case of Sy1 than in Sy2. 
Finally, data from 200 to 500 $\mu$m are found to be useful, even if not as much as data in the 
10-200 $\mu$m range, for constraining the torus model parameters. It is important to note that, for simplicity, 
here we are not considering the angular resolution provided by the different instruments. 
Thus, aside from the measured expected utility of a given filter, it is important to consider the resolution 
of the data when deciding the next observation.

We performed a BAE experiment for the galaxy NGC 3081 and found
that having two data points only --one in the NIR and another in the MIR--  significantly 
constrains the the torus width ($\sigma$), which does not change when adding further IR data points. 
The addition of FIR data helps constraining the torus radial extent ($Y$) and the number of 
clouds (N$_0$). Finally, the combination of the J, Si2 and Q filters
appears to be the most suitable to constrain the inclination angle of the torus ($i$). 
Note, however, that these
are the results found for a particular galaxy, and they may not be applicable to all Seyferts.

Given the large pressure on large-aperture telescopes, our approach --when combined with spatial
resolution considerations-- is very promising for
constraining CLUMPY models, and possibly torus models in general, in sparsely sampled SEDs with the smallest number
of observed points. 
The Bayesian adaptive exploration scheme can be applied to any
parametric model (e.g., smooth torus, light curves of classical eclipsing binaries, 
x-ray binaries) once the tools to carry out a fully Bayesian inference are available. The application
to the CLUMPY models is straightforward thanks to the \B\ tool. We note that the computation of the
expected utility is already included in the public version of \B, which is available for the community.

\section*{acknowledgements}
AAR and CRA acknowledge Almudena Alonso Herrero for very useful comments.
AAR acknowledges financial support by the Spanish Ministry of Economy and Competitiveness through 
projects AYA2010-18029 (Solar Magnetism and Astrophysical Spectropolarimetry) and 
Consolider-Ingenio 2010 CSD2009-00038. CRA acknowledges the Spanish Ministry of Science and 
Innovation (MICINN) through project Consolider-Ingenio 2010 Program grant CSD2006-00070: First Science
with the GTC (http://www.iac.es/consolider-ingenio-gtc/) and the Spanish Plan Nacional grant 
ESTALLIDOS AYA2010-21887-C04.04. We finally acknowledge useful comments from the anonymous referee.


\label{lastpage}

\end{document}